# Quantum frequency doubling in the topological insulator $Bi_2Se_3$

Pan He[1,3,†], Hiroki Isobe[2,†], Dapeng Zhu[1], Chuang-Han Hsu[1], Liang Fu[2,*], and Hyunsoo Yang[1,*]

[1]*Department of Electrical and Computer Engineering, National University of Singapore, 117576, Singapore*

[2]*Department of Physics, Massachusetts Institute of Technology, Cambridge, Massachusetts 02139, USA*

[3]*Institute for Nanoelectronic devices and Quantum computing, Fudan University, Shanghai 200433, China*

[†]These authors contributed equally to this work. [*]e-mail: liangfu@mit.edu; eleyang@nus.edu.sg

**Abstract**

**The nonlinear Hall effect due to Berry curvature dipole (BCD) induces frequency doubling, which was recently observed in time-reversal-invariant materials. Here we report novel electric frequency doubling in the absence of BCD on a surface of the topological insulator $Bi_2Se_3$ under zero magnetic field. We observe that the frequency-doubling voltage transverse to the applied ac current shows a threefold rotational symmetry, whereas it forbids BCD. One of the mechanisms compatible with the symmetry is skew scattering, arising from the inherent chirality of the topological surface state. We introduce the Berry curvature triple, a high-order moment of the Berry curvature, to explain skew scattering under the threefold rotational symmetry. Our work paves the way to obtain a giant second-order nonlinear electric effect in high mobility quantum materials, as the skew scattering surpasses other mechanisms in the clean limit.**

## Introduction

The Hall effect, the generation of voltage transverse to an electric current and a magnetic field, and the anomalous Hall effect (AHE) in magnetic materials[1] require time-reversal symmetry breaking. These effects refer to a transverse electric response in the linear region, where the Hall voltage $V_y$ scales linearly with the longitudinal current $I_x$. The second-order (nonlinear) Hall effect, in which $V_y$ depends quadratically on $I_x$, has attracted attention in condensed matter physics[2-4]. A quantum origin of the nonlinear Hall effect in time-reversal-invariant materials is the Berry curvature dipole (BCD)[3]. The nonlinear Hall effect due to the BCD was observed recently in bilayer and few-layer $WTe_2$[5,6]. The BCD generates an effective magnetic field in a stationary state, thus leading to the nonlinear Hall effect[3]. Electrical second-harmonic generation (SHG), including the nonlinear Hall effect, can exist only when a system lacks inversion symmetry[7-9]. Despite growing interest of BCD[10-14], it is subject to strict crystal symmetry restrictions and vanishes in certain crystals even without inversion symmetry[3], while second-order response is still allowed. Therefore, a search for electrical SHG independent of the BCD is desirable.

Inversion symmetry is absent in low-symmetry crystals (such as $WTe_2$[5,6,10]), and on a surface or an interface. However, the electrical SHG has not explored in surface/interface systems with time-reversal symmetry. Three-dimensional (3D) topological insulators (TIs) have attracted great interest due to the topological surface state (TSS) with spin-momentum locking[15-17] for applications in spintronics and quantum computing[18-20]. With an inversion-symmetric bulk, 3D TIs such as $Bi_2Se_3$, $Bi_2Te_3$, and $Sb_2Te_3$ host electrical SHG only on the surfaces. Furthermore, threefold rotational symmetry of the TI surface in Fig. 1a forces a BCD to vanish (Fig. 1b)[3]; thus, the BCD-induced nonlinear Hall effect is not allowed. In addition to the intrinsic contribution by a BCD, extrinsic effects arising from impurity or phonon scatterings, as intensively studied in AHE[1], are yet to be well sorted out for nonlinear effects. 3D TIs are ideal platforms in searching for extrinsic electrical SHG in the absence of a BCD. While recent theoretical studies addressed extrinsic mechanisms[21-24], an experimental observation of extrinsic contributions to the electrical SHG has not been reported.

In this work, we show the observation of electrical SHG in the 3D TI $Bi_2Se_3$ with time-reversal symmetry. The transverse voltage response depends quadratically on the applied current in the nonmagnetic $Bi_2Se_3$ films under zero magnetic field. The observed second-order response follows a threefold rotational symmetry on the surface of $Bi_2Se_3$. Notably, the symmetry excludes a BCD, which distinguishes the mechanism for electrical SHG from the previous studies[5,6]. We consider our observation arising dominantly from skew scattering in the TSS with its inherently chiral wave function. Instead of a BCD, we introduce the Berry curvature triple, which quantifies the moment of the Berry curvature under the threefold rotational symmetry. The skew scattering mechanism applies to a much wider class of noncentrosymmetric materials as broken inversion is the only symmetry constraint unlike the BCD.

## Results

**Observation of electric SHG**

High-quality $Bi_2Se_3$ films were grown on $Al_2O_3$ (0001) substrates in a molecular beam epitaxy system. The first quintuple layer (QL) of $Bi_2Se_3$ is completely relaxed by van der Waals bonds[25]. In addition, the lattice constant of $Bi_2Se_3$ film relaxes to its bulk value, implying the absence of strain from the substrate[25]. Thus, the induction of BCD via breaking the threefold rotational symmetry[26] does not occur in $Bi_2Se_3$ films, as confirmed by our angle dependent transport measurements below. Multiple Hall bar devices with current channels along different crystalline directions (Fig. 1c) were fabricated. Figures 1d,1e show the basic electrical characterization. The longitudinal resistivity $\rho$ (Fig. 1d) shows a typical metallic behavior and saturates below ~30 K[27,28]. Figure 1e displays the longitudinal $R_{xx}$ and Hall $R_{yx}$ resistances as a function of an out-of-plane magnetic field at 2 K. $R_{xx}$ at the low field region exhibits the effect of weak anti-localization, indicative of 2D surface transports[29]. $R_{yx}$ depends linearly on the magnetic field, from which the $n$-type carrier density $n_{2D}$ is extracted to be ~$6.26\times10^{13}$ cm$^{-2}$. $n_{2D}$ changes less than 2.3% for temperature ($T$) of $2 < T < 300$ K.

To explore the nonlinear electric transport, we perform harmonic measurements using low-frequency lock-in techniques schematically shown in Fig. 2a. We apply the ac current $I_x(t) = I \sin \omega t$ along the $x$ direction and measure the voltage $V_y$ perpendicular to the current. Under time-reversal and threefold rotational symmetries, the transverse voltage response does not contain the linear contribution, leading to the expression

$$V_y = R^{(2)}_{yxx} I_x^2, \tag{1}$$

which contains the SHG signal $V_y^{2\omega} = \frac{1}{2} R^{(2)}_{yxx} I^2 \sin(2\omega t - \pi/2)$. Note that the coefficient $R^{(2)}_{yxx}$ is proportional to the second-order conductivity $\sigma^{(2)}_{yxx}$ (see Supplementary Information S1), which can be finite in noncentrosymmetric materials[3].

Figure. 2b shows the second harmonic transverse voltage under zero magnetic field in 20 QL $Bi_2Se_3$. Its quadratic dependence on the ac current ($V_y^{2\omega} \propto I^2$) reveals the electrical SHG from a time-reversal-invariant 3D TI. Equivalently, the second harmonic transverse resistance defined as $R_{yx}^{2\omega} \equiv V_y^{2\omega}/I$ scales linearly with *I* (Fig. 2c). Moreover, it changes the sign when we invert the current direction and the corresponding Hall probes (schematic in the inset of Fig. 2c). This is consistent with the second-order nature of nonlinear transport in Eq. (1). The electric SHG has little dependence on the input frequencies ranging from 9 to 263 Hz (see Supplementary Information S2).

Figure 2d displays the $R_{yx}^{2\omega}(I)$ data at different temperatures. The slope of $R_{yx}^{2\omega}(I)$ (i.e. $R_{yxx}^{(2)}$) quantifies the magnitude of the electrical SHG. $R_{yxx}^{(2)}$ decreases gradually as temperature increases in Fig. 2e. In general, finite temperature affects the nonlinear electric transport through thermal smearing of the electron distribution function *f* and the change of the electron scattering time *τ*. Thermal smearing has little effect on the result as the Fermi energy is much higher than thermal energy $k_B T$ in our Bi$_2$Se$_3$ ($k_B$: the Boltzmann constant). To reveal the effect of *τ*, we depict the measured carrier mobility *μ* in Fig. 2f. Both the SHG signal and mobility tend to decrease as temperature rises.

**Angular dependence and scaling of nonlinear transport**

To characterize the angular dependence of nonlinear electric transport, we measure various devices with the current applied along different crystal directions on 20 QL Bi$_2$Se$_3$ (Fig. 1c). The current direction is denoted by angle Θ with respect to the $\overline{\Gamma K}$ direction (i.e., [-1, 1, 0] direction on Bi$_2$Se$_3$ (111) surface of the primitive lattice in real space) in Fig. 3. $R_{yx}^{2\omega}$ shows the maximum value when the current direction is along $\overline{\Gamma K}$ (Figs. 2b,2c), and decreases when the current is rotated 15° away from $\overline{\Gamma K}$ in Fig. 3a. For Θ = 30°, i.e., with the ac current along the $\overline{\Gamma M}$ direction, $R_{yx}^{2\omega}$ becomes vanishingly small (Fig.

3b). $R_{yx}^{2\omega}$ switches sign with a similar magnitude when the current direction is rotated by 60° from the $\overline{\Gamma K}$ to $\overline{\Gamma K'}$ direction in Fig. 3c. The small non-symmetry of $R_{yx}^{2\omega}(I)$ at the positive and negative current in Figs. 3a-3c can be due to misalignments of Hall bar. The electric SHG measured at 24 different directions is summarized in Fig. 3d, which shows the threefold angular dependence of $R_{yxx}^{(2)}$. The similar angular dependence is also observed in 10 QL Bi$_2$Se$_3$ (Supplementary Information S3). We emphasize that threefold rotational symmetric signal with sign change excludes the Joule heating effect as an origin, which is isotropic and generally leads to the third harmonic generation. The threefold symmetry also excludes a BCD, while the helical spin texture[30] and the Berry curvature[31] (Fig. 1b) on the hexagonally warped Fermi surface (FS) of the TSS[32,33] share the same angular dependence. We note that the Berry curvature has the opposite sign along $\overline{\Gamma K}$ and $\overline{\Gamma K'}$ due to time-reversal symmetry.

The nontrivial wavefunction on the TSS with scattering by impurities or phonons can give rise to finite electrical SHG[24]. To investigate the microscopic mechanism, we examine the scaling properties of the second-order transport with respect to the linear conductivity $\sigma$ of the film using the data in Fig. 1d and Fig. 2e. Figure 4a shows that the experimental data fit well with $\frac{E_y^{(2)}}{E_x^2} = a\sigma^2 + b$, where $E_y^{(2)} = \frac{V_y^{2\omega}}{W}$ and $E_x = \frac{V_x^{\omega}}{L}$ ($W$ and $L$ are the width and length of the sample, respectively). The linear and second-order conductivities $\sigma$ and $\sigma_{yxx}^{(2)}$ are related by $J_y^{(2)} = \sigma_{yxx}^{(2)} E_x^2 = \sigma E_y^{(2)}$, so the coefficients $a$ and $b$ represent contributions in $\sigma_{yxx}^{(2)}$ that scale as $\sigma^3$ and $\sigma$, respectively. Furthermore, $\sigma$ is proportional to $\tau$ for low frequencies compared to $\tau^{-1}$. Therefore, the intercept $b$ amounts to the $\tau$ linear contributions of the second-order conductivity, which are generally attributed to BCD[3] and/or side jump[6]. Note that the former is absent in our case for the symmetry reason, so we attribute the $\tau$-linear contribution to side jump. On the other hand, the slope $a$ quantifies the contribution $\sigma^{(2)}_{yxx} \propto \tau^3$, which originates from skew scattering as we discuss below. We obtain similar fitting results for $\Theta = 15°$ in Fig. 4b and also in 10 QL $Bi_2Se_3$ (see Supplementary Information S3). Notably, the cubic contribution plays a dominant role over the linear one as $\sigma$ increases in $Bi_2Se_3$, and these two contributions are of opposite signs as shown in Fig. 4a and 4b and are separated in Supplementary Information S4. The scaling of electrical SHG with respect to the surface linear conductivity $\sigma_s$ is also analyzed in Supplementary Information S5.

**Physical origin of nonlinear transport**

The TI $Bi_2Se_3$ possesses time-reversal and inversion symmetries in the bulk. However, inversion is broken on the surface and hence the metallic TSS with $C_{3v}$ crystalline symmetry can host electrical SHG. It takes the form[24,34]

$$J = \sigma^{(2)} |\mathbf{E}|^2 \cos 3\Theta, \quad (2)$$

where $\Theta$ is the angle of the applied electric field $\mathbf{E}$ with respect to the $\overline{\Gamma K}$ direction and the current density $J$ is measured perpendicular to $\mathbf{E}$. There is only one independent element $\sigma^{(2)}$ in the second-order conductivity tensor $\sigma^{(2)}_{abc}$ for a two-dimensional system with $C_{3v}$ symmetry (see Methods).

Skew scattering is one of the microscopic mechanisms that contributes to $\sigma^{(2)}$. It arises even classically when there are nontrivial impurity potentials lacking inversion on the atomic scale[8,34,35] or by local correlation of spins[36]. Alternatively, without relying details of impurities, quantum Bloch functions can imprint inversion breaking and trigger skew scattering, which is the case for the TSS[24,34]. There is a semiclassical picture for skew scattering in a second-order process, schematically depicted in Fig. 4c. The hexagonally-warped Fermi surface consists of the positive and negative Berry curvature segments. Since both segments are anisotropic, they acquire finite but opposite velocities in the second-order response. When we construct a wave packet from states on the Fermi surface, it self-rotates due to finite Berry curvature and the rotation direction depends the sign of Berry curvature. Like the Magnus effect, even an isotropic scatterer deflects the motion of wave packets in a preferred direction due to the self-rotation, thus leading to finite SHG.

The semiclassical Boltzmann transport calculation[24] based on the model Hamiltonian of TSS[32,37] leads to the linear conductivity from the TSS $\sigma_{\text{TSS}} = \frac{e^2 \tau \epsilon_F}{4\pi \hbar^2}$ and the second-order conductivity from skew scattering is given by $\sigma^{(2)} = \frac{e^3 v \tau^3}{\hbar^2 \tilde{\tau}}$, where $\tau$ is the transport scattering time, $\tilde{\tau}$ is the skew scattering time, $e$ is the electric charge, $\epsilon_F$ is the Fermi energy, and $v$ is the Dirac velocity. Importantly, skew scattering yields $\sigma^{(2)} \propto \tau^3$ (assuming that $\tilde{\tau}$ is constant) while other contributions including side jump have weaker powers in $\tau$, which distinguishes the skew scattering contribution. The experimentally observed $\sigma^{(2)}_{yxx} \propto \sigma^3$ behavior is supported by the skew scattering mechanism, whose contribution is the largest in our observations.

The second-order conductivity obeys the surface crystalline symmetry to have the form $\sigma^{(2)}_{yxx} = \sigma^{(2)} \cos 3\Theta$, according to Eq. (2) (Fig. 4d), which is in agreement with our experiment. Instead of a BCD, the threefold rotational symmetry inspires us to define the *Berry curvature triple* $T$, a higher-order moment of the Berry curvature distribution in the momentum space. It quantifies the strength of the Berry curvature on the Fermi surface, respecting threefold rotation: $T(\epsilon_F) = 2\pi\hbar \int \frac{d^2k}{(2\pi)^2} \delta(\epsilon_F - \epsilon_\mathbf{k})\Omega_z(\mathbf{k}) \cos 3\theta_\mathbf{k}$ ($\theta_\mathbf{k}$: the angle measured from the $\overline{\Gamma K}$ line). For the TSS, we obtain $T(\epsilon_F) = \frac{\lambda \epsilon_F}{2\hbar^2 v^3}$. The Berry curvature triple is related to the skew scattering time $\tilde{\tau}$. When we consider unscreened Coulomb impurities with the strength characterized by the dimensionless parameter $\alpha = \frac{e^2 Q}{4\pi\varepsilon_0 \varepsilon \hbar v}$, where $Q$ is the impurity charge, $\varepsilon_0$ is the vacuum permittivity, and $\varepsilon$ is the dielectric constant, we find $\tilde{\tau} \approx 4\pi^2 n_i \alpha^3 v^2 T(\epsilon_F)$ (see Supplementary Information S6).

We now provide the theoretical estimate of the second-order response from skew scattering. Though the second-order response arises only on the surface, both 2D surface and bulk states contribute to $\sigma$. As the contribution from the TSS is approximately 40% from the top and bottom surfaces[38], we estimate $\tau \approx 0.1$ ps and $\tilde{\tau} \approx 10$ ps (see Methods). The ratio $\tau/\tilde{\tau}$ of ~ 1% quantifies the relative strength of skew scattering. The estimated $\tau$ and $\tilde{\tau}$ result in the theoretical value $\sigma^{(2)} \approx 1.0 \times 10^{-11}$ A·V$^{-2}$·m. This is about three times larger than the experimentally observed value $\sigma^{(2)} = 2.9 \times 10^{-12}$ A·V$^{-2}$·m. We can attribute this difference to the partial cancellation of the second-order response; the contribution of the top surface is dominant over that of the bottom surface. In addition, screening of the Coulomb interaction reduces the response (see Supplementary Information S6).

**Discussion**

We have demonstrated the electric SHG in a nonmagnetic 3D TI under zero magnetic field. It provides an example of BCD-independent nonlinear transverse transport, which is further revealed to arise from skew scattering. This skew scattering mechanism can be applicable to a broader class of noncentrosymmetric quantum materials, utilizing the chirality of electron wavefunction in Weyl and Dirac fermions[39]. Though our work reveals the nonlinear transport under low frequencies, it can be extended to higher frequency regimes such as GHz and THz. Thus, the electric SHG is complementary to previous optoelectronic approaches[34,40] to reveal the underlying physics of nonlinear effects.

Berry curvature is allowed to exist in the TSS[31,41], and concentrates in regions around $\bar{K}$ ($\bar{K}'$) points in Fig. 1b, leading to finite Berry curvature triple. Finite Berry curvature also affects the electron distribution function through the collision integral and the anomalous and side jump velocities[24]. The intrinsic contribution due to the anomalous velocity and hence BCD is absent in $Bi_2Se_3$ due to the symmetry reason[3]; however, the extrinsic contributions such as skew scattering and side jump persist[21]. The skew scattering contribution dominates in the weak impurity limit ($\tau \to \infty$)[23,24] because of its high-order $\tau$ dependence. Though a full quantitative understanding of various contributions to nonlinear electric transports remains elusive[21] which may include phonons, domain boundaries, impurities, and Berry curvature[42], identifying major mechanisms is an important step not only for the fundamental understanding of underlying principle, but for the development of rectification or second-harmonic devices for energy harvesting and high-frequency communication. The extrinsic nonlinear effect observed in $Bi_2Se_3$ is comparable in magnitude to the intrinsic one in few-layer $WTe_2$[6], which has a 2D nonlinear conductivity of $\sim 10^{-12}$ A·V$^{-2}$·m. Moreover, the extrinsic mechanism exemplified here applies to a wider class of materials with inversion-symmetry breaking, such as graphene/hexagonal-boron-nitride heterostructures[43], Dirac semimetal $ZrTe_5$[44,45] and the two-dimensional electron gas at the $LaAlO_3/SrTiO_3$ interface[46]. Engineering scattering processes in above materials is a promising way to achieve a prominent SHG by utilizing their much higher carrier mobilities. A higher mobility and long scattering time improve the efficiency in device applications since skew scattering has a higher order dependence on $\tau$[1,24,47].

**Methods**

**Sample preparation and electric measurements.** $Bi_2Se_3$ films were grown on $Al_2O_3$ (0001) substrates in a molecular beam epitaxy system with a base pressure $< 2\times10^{-9}$ mbar, as detailed in ref. [47]. Van der Waals epitaxy of $Bi_2Se_3$ film was achieved by adopting the two-step growth method[25,27,48,49]. For transport measurements, a capping layer of MgO (2 nm)/$Al_2O_3$ (3 nm) was deposited on top of the films prior to device fabrication. Hall bar devices were fabricated using the standard photolithography and Argon plasma etching. They were wire-bonded to the sample holder and installed in a physical property measurement system (PPMS, Quantum Design) for transport measurements. We performed low-frequency ac harmonic electric measurements, using Keithley 6221 current sources and Stanford Research SR830 lock-in amplifiers. During the measurements, a sinusoidal current with a constant amplitude and certain frequency is applied to the devices, and the in-phase first harmonic $V_\omega$ and out-of-phase second harmonic $V_{2\omega}$ longitudinal and transverse voltages were measured simultaneously by four lock-in amplifiers.

**Theoretical modeling and estimate.** The Hamiltonian for the TSS is[32,37]

$$H = \hbar v(k_x \sigma_y - k_y \sigma_x) + \frac{\lambda}{2}(k_+^3 + k_-^3)\sigma_z, \tag{3}$$

where $k_\pm = k_x \pm i k_y$, $\sigma_a$ denotes the Pauli matrix ($a = x, y, z$), and $\lambda$ quantifies the hexagonal warping[32]. In this section, the $x$ axis is set perpendicular to the reflection plane, i.e., along the $\bar{\Gamma}\bar{K}$ line. For the surface state of $Bi_2Se_3$, we find $v = 5 \times 10^5$ m/s and $\lambda = 80$ eV $\cdot$ Å$^3$, and the FS is located above the Dirac point, where a hexagonally warped FS was found[30,33].

In general, the current response quadratic to the electric field $E$ takes the form $J_a^{(2)} = \sigma_{abc}^{(2)} E_b E_c$, where $\sigma_{abc}^{(2)}$ is the second-order conductivity. For a two-dimensional system with $C_{3v}$ symmetry like the TSS, it has only one independent element $\sigma^{(2)} \equiv \sigma_{xxy}^{(2)} = \sigma_{xyx}^{(2)} = \sigma_{yxx}^{(2)} = -\sigma_{yyy}^{(2)}$. To estimate the transport properties, we assume Coulomb impurities, randomly distributed in a sample. Taking account of the Thomas-Fermi screening, we write the Fourier transform of the Coulomb interaction as $V(q) = \frac{2\pi\alpha\hbar v}{q+q_{TF}}$, where $q_{TF}$ is the Thomas-Fermi wavevector. Here, we consider unscreened Coulomb impurities ($q_{TF} = 0$), which we discuss below.

In estimating $\tau$ and $\tilde{\tau}$, we use the dielectric constant[50] $\epsilon \approx 100$, leading to $\alpha \approx \frac{1}{23}$. We use the previous observation that the contribution of the TSS from the top and bottom surfaces to the total conduction is about 40%[38] and assume that the impurity density $n_i$ is approximately the same as the carrier density $n_{2D}$. Thus, the observed linear conductivity $\sigma = 2.5 \times 10^{-3}$ $\Omega^{-1}$ at 10 K leads to the carrier density of the TSS $n_{TSS} = 2.43 \times 10^{12}$ cm$^{-2}$, the corresponding Fermi wavelength $\lambda_F = \sqrt{\frac{\pi}{n_{TSS}}} = 11.4$ nm, the scattering time $\tau \approx 0.1$ ps, and the skew scattering time $\tilde{\tau} \approx 10$ ps, where we use the expressions[24] $\sigma_{\text{TSS}} = \frac{e^2 \tau \epsilon_F}{4\pi \hbar^2}$, $\tau^{-1} = \frac{\pi}{2} n_i \alpha^2 \nu \lambda_F$, and $\tilde{\tau}^{-1} = \frac{4\pi^3}{\hbar} \frac{n_i \alpha^3 \lambda}{\lambda_F}$. The small ratio of $\frac{\tau}{\tilde{\tau}} \ll 1$ satisfies the condition of the perturbative treatment of impurities in the semiclassical Boltzmann theory.

The Thomas-Fermi wavelength $\lambda_{\text{TF}} = \frac{2\pi}{q_{\text{TF}}}$ is typically ranging from 26 to 90 nm[51,52], resulting in the ratio $\lambda_F/\lambda_{\text{TF}} \lesssim 0.4$. We describe the detailed calculations and discussion about the effect of screening in Supplementary Information S6. We note that for short-range impurities or in the strong screening limit, i.e., $\lambda_{\text{TF}} \to 0$, skew scattering vanishes in a gapless Dirac system[24,34].

**Data availability.** The data that support the plots within this paper and other findings of this study are available from the corresponding author upon reasonable request.

**Acknowledgments**

We thank Steven S.-L. Zhang, and G. Vignale for discussions. The work was partially supported by SpOT-LITE program (A*STAR grant, A18A6b0057) through RIE2020 funds, and Singapore Ministry of Education (MOE) Tier 1 (R263-000-D61-400-114). P. He acknowledges the start-up funding from Fudan University.

**Author contributions**

P.H. fabricated the devices, performed transport measurements and analyzed the data. D.Z. grew the films. H.I. and L.F. performed theoretical studies. C.-H.H. contributed to Berry curvature calculation. All authors discussed the results. P.H., H.I., L.F., and H.Y. wrote the manuscript.

**Additional information**

Supplementary information is available in the online version of the paper. Reprints and permissions information is available online at www.nature.com/reprints. Correspondence and requests for experimental materials should be addressed to H.Y. Correspondence and requests for theoretical materials should be addressed to L.F.

**Competing interests**

The authors declare no competing interests.

**Fig. 1. Crystal structure, Berry curvature distribution and characterizations of Bi$_2$Se$_3$.**
**a,** Crystal structure of Bi$_2$Se$_3$. From top view along the $z$-direction of Bi$_2$Se$_3$ (111) surface, the triangle lattice in one quintuple layer has three different positions, denoted as Se1, Bi and Se2. **b,** Schematic of Berry curvature distribution on the Fermi surface (FS) of TSS. The colored hexagon represents the hexagonally warped FS. The blue and red color contours indicate the negative and positive Berry curvature in an arbitrary unit. The black hexagon represents the surface Brillouin zone. **c,** Schematic of the sample structure and optical image of Hall bar devices with the current channel along different directions. The inset shows magnified view of a device. The scale bar is 100 μm. **d,** Temperature dependence of the longitudinal resistivity ($\rho_{xx}$) under zero magnetic field in 20 QL Bi$_2$Se$_3$. **e**, Magnetic field dependence of the longitudinal resistance ($R_{xx}$) and Hall resistance ($R_{yx}$) at 2 K in 20 QL Bi$_2$Se$_3$.

**Fig. 2. Observation of electric SHG under zero magnetic field in Bi$_2$Se$_3$. a**, Schematic illustration of the nonlinear transport measurements in a Hall bar device using the second harmonic method. **b**, The second harmonic transverse voltage $V_y^{2\omega}$ versus ac current amplitude $I$ in 20 QL Bi$_2$Se$_3$ at 50 K. The solid line is a quadratic fit to the data. Blue hexagon in the inset represent the FS of Bi$_2$Se$_3$. Red line is along the $\overline{\Gamma K}$ direction, and the black arrow denotes the current direction in **k**-space. **c**, The second harmonic transverse resistance $R_{yx}^{2\omega}$ scales linearly with $I$ at 50 K. It changes sign with reversing the current direction and corresponding Hall probes. **d**, The $R_{yx}^{2\omega}(I)$ curves measured at different temperatures from 20 to 200 K. **e**, The slope of $R_{yx}^{2\omega}(I)$ curves ($R_{yxx}^{(2)}$) as a function of temperature. Error bars correspond to the standard error of linear fitting. **f**, The measured carrier mobility µ as a function of temperature.

**Fig. 3. Angular dependence of nonlinear transport in Bi$_2$Se$_3$.** The second-harmonic transverse resistance versus current for three typical current injection angles at $\Theta = 15°$ (**a**), 30° (**b**), and 60° (**c**). Blue hexagon in the inset represents the FS of Bi$_2$Se$_3$. When current direction reverses, the second harmonic resistance changes sign in all current directions. The solid lines are linear fits to the data. **d**, The slope of $R_{yx}^{2\omega}(I)$ curves ($R_{yxx}^{(2)}$) as a function of current direction. The data in panels **a-d** were collected at $T = 50$ K.

**Fig. 4. Physical origin of nonlinear transport. a,b** $E_y^{(2)}/E_x^2$ versus the square of the longitudinal conductivity ($\sigma^2$) for the current injection angles at $\Theta = 0°$ (**a**), 15° (**b**) in temperature of 50-200 K, where the electron transport is strongly determined by electron-acoustic phonon scattering[28]. The red line is a linear fit to the experimental data. **c**, Schematic plot of the skew scattering of surface Dirac fermion and the induced nonlinear transverse transport. **d**, The theoretical nonlinear transverse conductivity versus the applied electric field direction $\Theta$ based on the skew scattering of TSS.

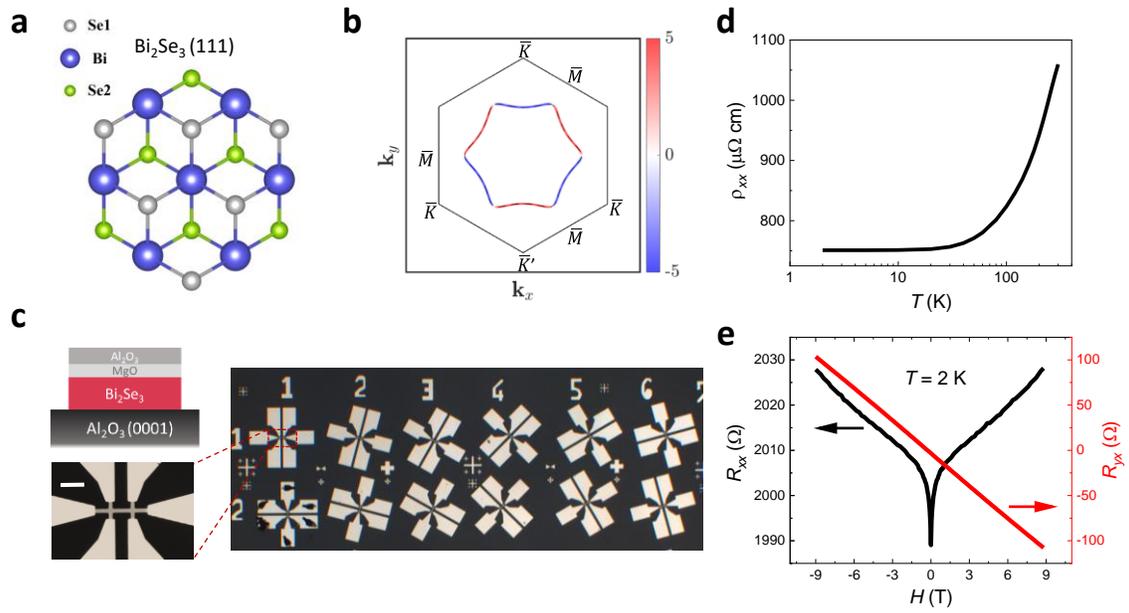

Figure 1

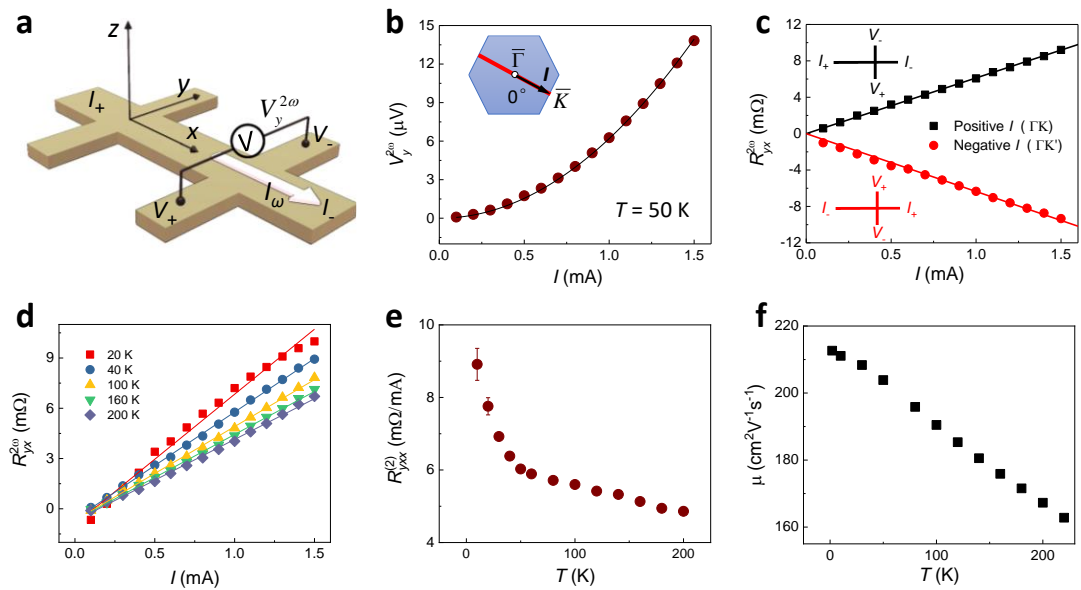

Figure 2

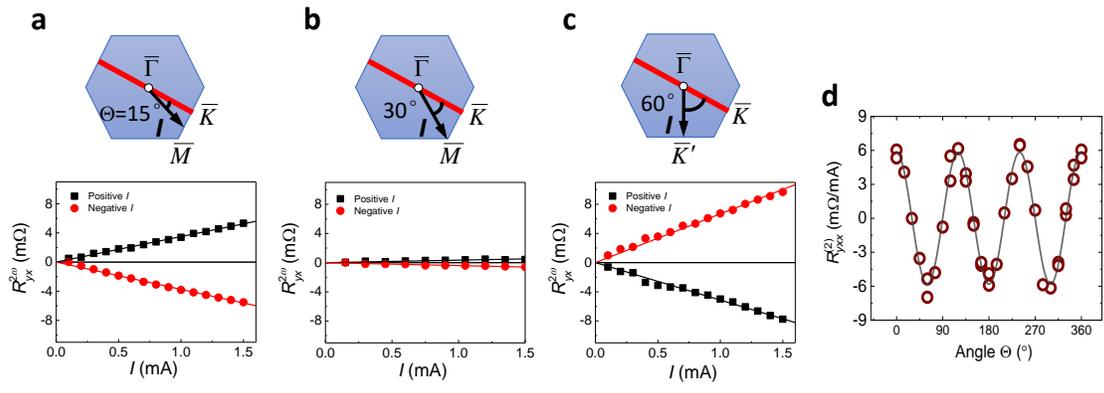

Figure 3

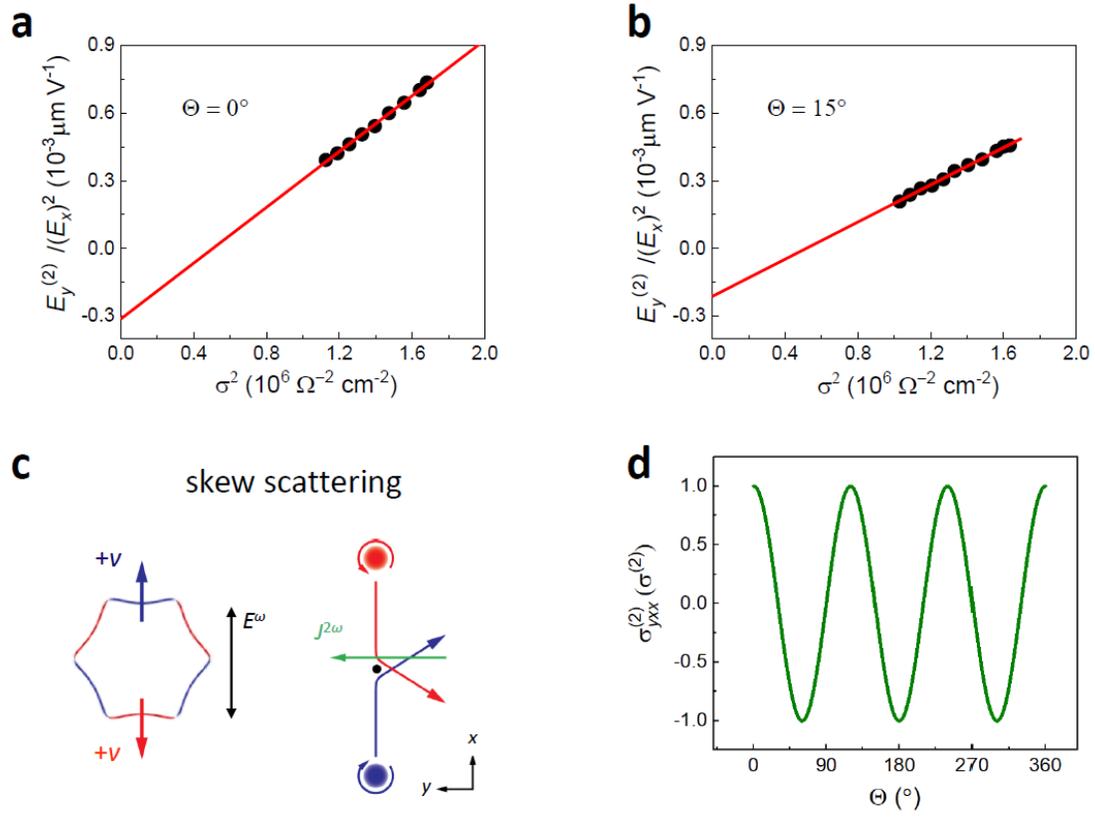

Figure 4